\documentclass[11pt]{elsart}
\usepackage[dvips]{graphicx}
\usepackage{amsmath}
\usepackage{amssymb}
\usepackage[english]{babel}

\journal{Physica A}

\begin{document}
\begin{frontmatter}

\title{Vibrational Features of Water at the
Low-Density/High-Density Liquid Structural Transformations}

\author[label1]{Ramil M. Khusnutdinoff}
\ead{khrm@mail.ru}
\author[label1]{Anatolii V. Mokshin}
\ead{anatolii.mokshin@mail.ru}
\address[label1]{Department of Physics, Kazan (Volga region) Federal University,\\
Kremlevskaya Street 18, 420008 Kazan, Russia}

\begin{abstract}
A structural transformation in water upon compression was
recently observed at the temperature $T=277$~K
in the vicinity of the pressure $p \approx 2\;000$~Atm
[R.M. Khusnutdinoff, A.V. Mokshin, J. Non-Cryst. Solids \textbf{357}, 1677 (2011)].
It was found that the transformations are related with the principal structural changes
within the first two coordination shells as well as the deformation of
the hydrogen-bond network. In this work we study in details the
influence of these structural transformations on the vibrational molecular dynamics of water
by means of molecular dynamics simulations on the basis of the model Amoeba
potential ($T=290$~K, $p=1.0 \div 10\;000$~Atm). The equation of state and the isothermal
compressibility are found for the considered ($p$,$T$)-range.
The vibrational density of states extracted for $THz$-frequency range manifests the two distinct
modes, where the high-frequency mode is independent on pressure
whereas the low-frequency one has the strong, non-monotonic
pressure-dependence and exhibits a step-like behavior at the pressure $p
\approx 2000$~Atm. The extended analysis of the local structural and vibrational
properties discovers that there is a strong correlation between the primary
structural and vibrational aspects of the liquid-liquid structural transformation
related with the molecular rearrangement within
the range of the second coordination shell.
\end{abstract}

\begin{keyword}
Amoeba water model \sep molecular dynamics
\sep equation of state
\sep local isothermal compressibility
\sep vibrational density of states
\PACS
61.20.Ja \sep 61.25.Em \sep 67.25.dt \sep 68.35.Rh
\end{keyword}

\date{\today}
\end{frontmatter}

\section{Introduction}

The physico-chemical features and soluble properties of water are of the central
importance for life on our planet.
However, despite of a knowledge of some properties
at the ambient conditions that corresponds only to narrow part of water phase diagram,
the inherent structural and dynamical properties are still far from
to be completely understood, that motivates to perform a large number
of experimental and theoretical investigations of water
\cite{Khusnutdinoff_2011,Soper_2008a,Soper_2000,Katayama_2010,Davitt_2010,Krisch_2002,
Jansson_2010,Liu_2002,Li_2010,Moore_2010,Soper_2008,Vega_2009,Paschek_2005}.
Water is known as a 'system' with a rich variety of anomalous
behavior related with the different phase states. For example, in a
solid state, water has at least fifteen crystalline forms, where four of
these can coexist with the liquid phase, and a variety of the
amorphous phases known as amorphous ices and polyamorphism \cite{Soper_2008a,Salzmann_2009,Brovchenko_2008}.

\begin{figure}
\begin{center}
\includegraphics[width=1.0\textwidth]{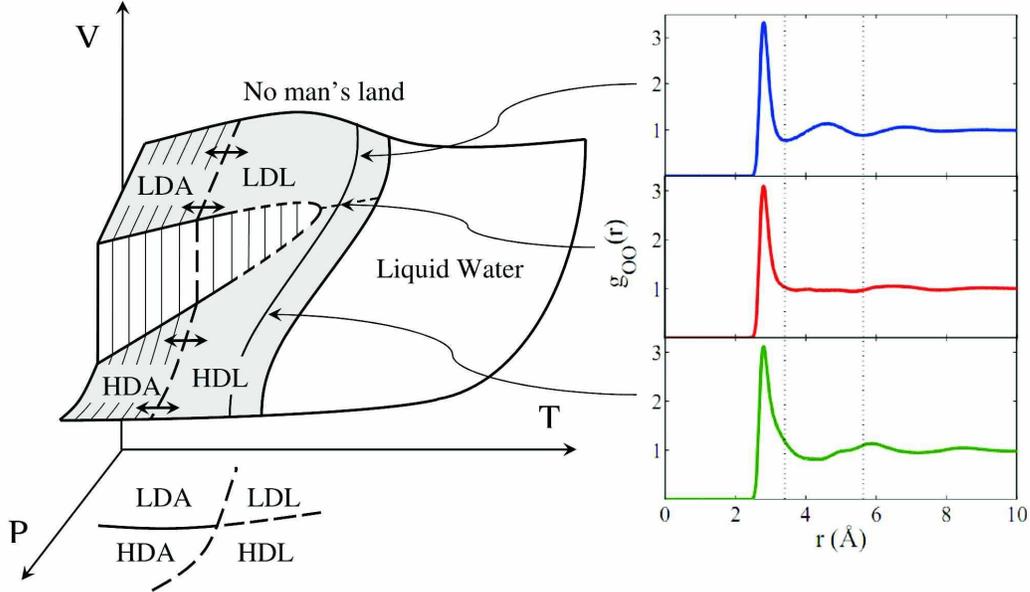}
\end{center}
\caption[kurzform]{\label{Fig_PhaseD} (Color online) Left panel: Schematic
phase diagram of liquid-amorphous water (see also Ref. \cite{Mishima_2010}).
White range corresponds to a stable liquid water, gray region indicates the
liquid-amorphous phase, which is characterized by the separated low-density
and high-density states. Broken line is the conventional boundary between
the liquid and the amorphous phases. Solid line is the 'isotherm' considered in the
work, where a sharp jump in density is no observed. LDL/HDL-structural
transformations arise at the temperature $T\in [260~\textrm{K}; 400~\textrm{K}]$
and the pressures $p\in [2000~ \textrm{Atm}; 4000~ \textrm{Atm}]$
(see Fig. 5 in Ref. \cite{Saitta_2003} and Fig. 7 of Ref. \cite{Li_2005}).
Right panel: Radial distribution functions $g_{OO}(r)$ typical for three
ranges of the phase diagram (LDL, LDL/HDL-boundary, HDL);
the correspondence is indicated by arrows.}
\end{figure}

One of the interesting features of water is related with the
first-order like phase transition from low-density liquid
(LDL) state to high-density liquid (HDL) state was reported in
Ref.~\cite{Poole_1992}.
The hypothesis of a ``liquid-liquid phase transition'' in water
was considered in a series of experimental, computational
and theoretical studies (see, for example, Refs.
\cite{Mishima_1998,Debenedetti_2003,Debenedetti_2003b} and references therein).
As it was found, the difference between these liquid states
is mainly provided by the structural properties. Namely, in the HDL,
the local tetrahedrally coordinated hydrogen bond (HB) structure is not
fully developed, whereas in the LDL, a more open, locally
``ice-like'' HB network is realized \cite{Khusnutdinoff_2011,Soper_2000}.
Both these liquid states are closely associated with
the two amorphous phases observed experimentally in water: the
low-density amorphous (LDA) and the high-density amorphous (HDA)
ices, although locus of this ``transition'' is still debatable
\cite{Brovchenko_2008,Debenedetti_2003,Debenedetti_2003b,Loerting_2006}.
Remarkably, that a sharp jump in density (volume) is smoothed with the
increase of temperature (or the decrease of pressure) and disappears
completely, while LDL/HDL structural transformations still exists. It is interesting
to note that the amorphous phase of water is also characterized by
the ``transition'' from HDA ice to very high-density amorphous (VHDA) ice
\cite{Loerting_2001,Finney_2002a,Finney_2002b,Loerting_PRL2006,Winkel_2008}.
A schematic phase diagram of water, that treats qualitatively some empirical results,
is given in Fig. \ref{Fig_PhaseD}.
Above the critical temperature of the predicted LDL/HDL transition, the jump in volume
(density) is absent, but the local structure changes completely with pressure. This is clear detected by the
radial distribution function (for example, for oxygen atoms), $g_{OO}(r)$,
where the first minimum disappears, while the second maximum and minimum
changes by a minimum and maximum, respectively (see right panel in Fig. \ref{Fig_PhaseD}).
It is significant to note that although a schematic phase diagram of
water is known in general sense for an extensive ($p$,$T$)-range (see, for example, Fig.~1 in
Ref.~\cite{Winkel_2008} and Fig.~4 in Ref.~\cite{Stanley_2007}), the
exact boundary location between LDL and HDL is not
well-defined as well as the smoothed LDL/HDL transformations range are still undetected.
So, the problem associated with the structural transformation and its
influence on microscopic dynamics of
liquid water (even at the room temperature) requires the additional
studies \cite{Brovchenko_2008}. The given work studies the structural features
within first two coordiantion shells \cite{Mokshin_2005} as well as the vibrational properties of liquid water at the
temperature $T=290$~K and a wide range of pressure by means of
molecular dynamics simulations on the basis of the Amoeba
interaction potential. The correlation between structural features
of the transformation and molecular vibrational dynamics of water is
also considered.
%%%%%%%%%%%%%%%%%%%%%%%%%%%%%%%%%%%%%%%%%%%%%%%%%%%%%%%%%%%
%% Figure  1
%%%%%%%%%%%%%%%%%%%%%%%%%%%%%%%%%%%%%%%%%%%%%%%%%%%%%%%%%%%
%%%%%%%%%%%%%%%%%%%%%%%%%%%%%%%%%%%%%%%%%%%%%%%%%%%%%%%%%%%%%%%%%

This paper organized as follows. The details of molecular dynamics
simulations of liquid water are presented in section 2.
The structural transformations within the second
coordination shell are explored by means of the local
isothermal compressibility $\kappa_T^0$ in section 3. The manifestation of the
effect of the structural transformations on the vibrational properties of water
and the possible correlation between these both are also
discussed here. Finally, we conclude with a short summary.

\section{Simulation Details \label{comp}}

Equilibrium classical molecular dynamics (MD) simulations of water
at the constant temperature $T=290$~K and the different pressures
were performed  on the basis of the Amoeba\footnote{Recently, it was shown
in Ref. \cite{Khusnutdinoff_2011}, that the results of molecular dynamics
simulations on the basis of Amoeba interaction potential for structural and
dynamical properties of water are in a good agreement with the results on
ab-initio molecular dynamics simulations as well as the experimental
data on neutron diffraction \cite{Soper_2000b} and inelastic X-ray scattering \cite{Krisch_2002}.
Note that the Amoeba model reveals for the isobar $P=1.0$~Atm the
maximum in density at the temperature $T=290$~K instead of the
well-known value $277$~K \cite{Ren_2003}.
Actually, one notices that the Amoeba potential does not reproduce the
experimental equation of state correctly: for instance, the density
$1.21$ $g/cm^3$ at $290$~K is reached at $5000$~Atm in the present
simulation, instead of $8000$~Atm in the experiment.}
(Atomic Multipole Optimized Energetics for Biomolecular Applications) water model
suggested recently by Ren and Ponder \cite{Ren_2003,Ren_2004}. The
model uses a polarizable atomic multipole description of
electrostatic interactions, where the polarization is treated due to
the self-consistent induced atomic dipoles \cite{Ren_2003}, and a
modified version of the Thole's interaction model is used to damp
the induction at a short range \cite{Thole_1981,Ponder_2003}. Our
computations were performed for $4000$ water molecules interacted
within a cubic box with the periodic boundary conditions in all
directions. The Ewald summation was used to handle the electrostatic
interactions. Moreover, an atom-based switching window of the size
12~\textrm{\AA} was applied to cut off the van der Waals
interactions. The equations of motion were integrated via a modified
Beeman algorithm \cite{Allen_Tildesley},
\begin{eqnarray}
\vec{r}(t+\Delta\tau)=\vec{r}(t)+\vec{\vartheta}\Delta\tau+\frac{1}{6}\Bigg( 4\vec{a}(t)-\vec{a}(t-\Delta\tau) \Bigg) \Delta\tau^2 +\mathcal{O}(\Delta\tau^4), \nonumber \\
\vec{\vartheta}(t+\Delta\tau)=\vec{\vartheta}(t)+\frac{1}{12}\Bigg( 5\vec{a}(t+\Delta\tau)+8\vec{a}(t)-\vec{a}(t-\Delta\tau) \Bigg)
\Delta\tau+\mathcal{O}(\Delta\tau^3), \nonumber
\end{eqnarray}
with the time step $\Delta \tau = 1.0$~fs.
The isothermal-isobaric ensemble at the temperature $T=290$~K and the
pressure $p=1.0$, $1000$, $2000$, $2500$, $3000$, $3750$, $5000$, $7797$ and
$10\;000$~ Atm was applied by means of the Berendsen thermostat and barostat~\cite{Berendsen_1984}.
According to this scheme, to enforce the constant temperature
the system was weakly coupled to a heat bath with some temperature.
The velocities were scaled at an each step by such a way, that the rate of
the temperature change is
\begin{equation}
\frac{dT}{dt}=\frac{T_0-T}{\tau}.
\end{equation}
where $\tau$ is the coupling parameter, which determines how tightly the bath and the system are
coupled together, $T_0$ is the temperature of the external heat bath.
This method provides an exponential decay of the system temperature towards the desired
value. The change in temperature between the successive time steps is
\begin{equation}
\Delta T=\frac{\Delta\tau}{\tau}\bigg( T_0-T(t) \bigg),
\end{equation}
here
\begin{equation}
T(t)=\frac{1}{(3N-N_c)k_B}\sum_{i=1}^{N}\frac{\vec{p_i}^2}{m_i}
\end{equation}
the instantaneous value of the temperature; $N_c$ is the number of constraints
and $(3N-N_c)$ is the total number of degrees of freedom.
Thus, the scaling factor for the velocities is
\begin{equation}
\lambda=\sqrt{1+\frac{\Delta\tau}{\tau}\Bigg(\frac{T_0}{T(t)-1} \Bigg)}.
\end{equation}
Within the Berendsen barostat the system is made to obey the equation of motion
\begin{equation}
\frac{dp}{dt}=\frac{p_0-p}{\tau},
\end{equation}
where $p_0$ is the value of the applied pressure, $p$ is the instantaneous system pressure.
In the isotropic case, the box volume is scaled by a
factor $\eta$, whereas the coordinates are scaled by $\eta^{1/3}$, where
\begin{equation}
\eta=1-\frac{\beta \Delta \tau}{\tau}(p_0-p)
\end{equation}
and $\beta$ is the isothermal compressibility of the system. The value of the coupling parameter
$\tau$ in our simulations was equal to 0.1 ps.

\section{Results}
\subsection{Equation of State and Local Structural Properties \label{sate_equat}}

\begin{figure}
\begin{center}
\includegraphics[width=1.2\textwidth]{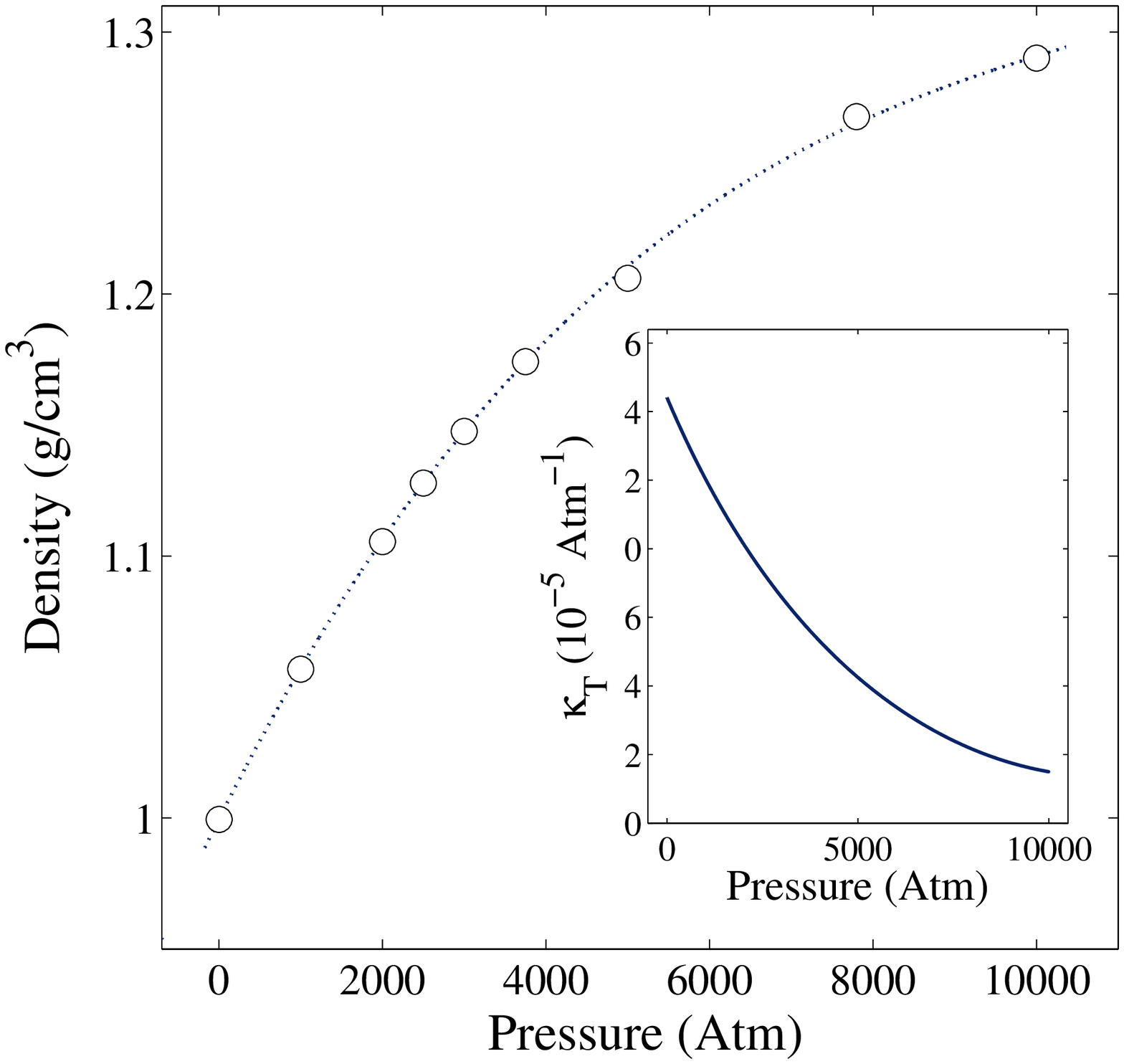}
\end{center}
\caption[kurzform]{\label{Fig_EOS} (Color online) Main: Equation of
state for liquid water at the temperature $T=290$~K. Inset:
Pressure dependence of the isothermal compressibility $\kappa_T$.}
\end{figure}

A central place at the study of any phase transition or structural
transformations takes the consideration of the equation of states.
Here, the equation of state was defined for the considered water model at
the fixed temperature $T=290$~K and different pressures. Results are shown in Fig. \ref{Fig_EOS}.
As can be seen from the figure, the presented
isotherm  $\rho(p)$ is well fitted by the cubic polynomial
\begin{equation}
\rho(p)=c_0+c_1p+c_2p^2+c_3p^3, \label{Eq_EOS}
\end{equation}
the values of the parameters $c_0$, $c_1$, $c_2$, $c_3$
are given in Table 1. The obtained equation allows
one to find directly the isothermal compressibility $\kappa_T$ as
\begin{equation}
\kappa_T=\frac{1}{\rho}\Bigg(\frac{\partial \rho}{\partial p} \Bigg)_T.
\label{Eq_CIC}
\end{equation}

There is no any visible singularity in the density $\rho(p)$ as well as in
the isothermal compressibility $\kappa_T(p)$ (see inset of Fig. \ref{Fig_EOS})
in the vicinity of the pressure $p_c \simeq 2000$~Atm, where LDL/HDL-structural
transformations were observed \cite{Khusnutdinoff_2011}.
This is direct evidence of the absence of the first-order-like
transition at LDL/HDL structural transformations, while in amorphous ices
this transition exists \cite{Mishima_2010}.

%%%%%%%%%%%%%%%%%%%%%%%%%%%%%%%%%%%%%%%%%%%%%%%%%%%%%%%%%%%
%% Figure  2
%%%%%%%%%%%%%%%%%%%%%%%%%%%%%%%%%%%%%%%%%%%%%%%%%%%%%%%%%%%
%%%%%%%%%%%%%%%%%%%%%%%%%%%%%%%%%%%%%%%%%%%%%%%%%%%%%%%%%%%%%%%%%
%%%%%%%%%%%%%%%%%%%%%%%%%%%%%%%%%%%%%%%%%%%%%%%%%%%%%%%%%%%%%%%%%

%%%%%%%%%%%%%%%%%%%%%%%%%%%%%%%%%%%%%%%%%%%%%%%%%%%%%%%%%%
% Table 1
%%%%%%%%%%%%%%%%%%%%%%%%%%%%%%%%%%%%%%%%%%%%%%%%%%%%%%%%%%
\begin{table}
\begin{center}
\caption{Parameters of equation (\ref{Eq_EOS}) of liquid water at
the temperature T=290~K.}
\begin{tabular}{l|c}
\hline
\hline
$~~~~~~~~~~~~~~~~~ \textrm{Coefficients}$ ~~~~~~~~~~~~~~  &
~~~~~~~~~~~~~~~~~~~~~~ 290~K ~~~~~~~~~~~~~~~~~~~~~~ \\
\hline
~~~~~~~~~~~~~~~~~$c_0$ ($\textrm{g/cm}^3$)                         & 0.9996 \\
~~~~~~~~~~~$c_1$ ($\textrm{g/cm}^3\cdot \textrm{Atm}^{-1}$)  & 6.204$\cdot 10^{-5}$  \\
~~~~~~~~~~~$c_2$ ($\textrm{g/cm}^3\cdot \textrm{Atm}^{-2}$)  & -4.642$\cdot 10^{-9}$ \\
~~~~~~~~~~~$c_3$ ($\textrm{g/cm}^3\cdot \textrm{Atm}^{-3}$)  & 1.349$\cdot 10^{-13}$ \\
\hline
\end{tabular}
\end{center}
\label{_tab1}
\end{table}
%%%%%%%%%%%%%%%%%%%%%%%%%%%%%%%%%%%%%%%%%%%%%%%%%%%%%%%%%%%%%%%%
%%%%%%%%%%%%%%%%%%%%%%%%%%%%%%%%%%%%%%%%%%%%%%%%%%%%%%%%%%%%%%%%

%%%%%%%%%%%%%%%%%%%%%%%%%%%%%%%%%%%%%%%%%%%%%%%%%%%%%%%%%%%
%% Figure  3
%%%%%%%%%%%%%%%%%%%%%%%%%%%%%%%%%%%%%%%%%%%%%%%%%%%%%%%%%%%
\begin{figure}
\begin{center}
\includegraphics[width=1.0\textwidth]{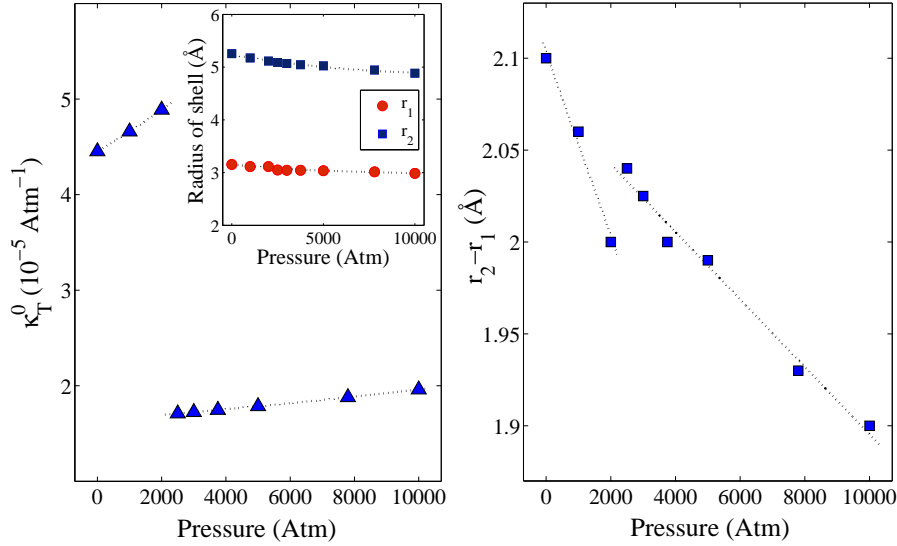}
\end{center}
\caption[kurzform]{\label{Fig_ISOLoc} (Color online) Main: Pressure
dependence of the local isothermal compressibility $\kappa_T^0$
(left panel) and the difference $(r_2-r_1)$ (right panel). Inset:
Radius of the first two coordination shells vs. pressure.}
\end{figure}
%%%%%%%%%%%%%%%%%%%%%%%%%%%%%%%%%%%%%%%%%%%%%%%%%%%%%%%%%%%%%%%%%
%%%%%%%%%%%%%%%%%%%%%%%%%%%%%%%%%%%%%%%%%%%%%%%%%%%%%%%%%%%%%%%%%
To perform a detailed study at microscopic level, we suggest to
consider the features upon the local spatial ranges around a water
molecule within the boundaries of the closest coordination shells.
For water the first two coordination numbers
defined as \cite{Yan_2007}
\begin{equation}
N(r_c)=4\pi n \int_0^{r_c} r^2 g(r) dr,
\label{Eq_CN}
\end{equation}
take the values $N(r_1)=4$ and $N(r_2)=16$, where  $r_1$ and $r_2$
is the radius of the first and the second coordination shell,
respectively; $n$ is the oxygen-number density. Therefore, on the
basis of Eq.~(\ref{Eq_CN}) the sizes of the first two coordination
shells $r_1$ and $r_2$ can be found for an arbitrary condensed phase
as the distances, where $N(r_1)=4$ and $N(r_2)=16$. Then, the local,
short-ranged properties can be extended to these of the spherical layer
enclosed between $r_1$ and $r_2$ of the volume $V_0=(4\pi/3)(
r_2^3-r_1^3)$. The isothermal compressibility extended to the
\emph{local} volume will be defined as
\begin{equation}
\kappa_T^0=-\frac{1}{V_0}\Bigg(\frac{\partial V_0}{\partial p} \Bigg)_T.
\label{Eq_CIC1}
\end{equation}
Fig. \ref{Fig_ISOLoc} presents the pressure dependence of these
characteristics: the local isothermal compressibility
$\kappa_T^0$; the radius of the first two coordination shells $r_1$ and $r_2$,
and the difference $(r_2 - r_1)$ associated with the linear size of the
considered local range. Contrary to the isothermal compressibility $\kappa_T$
(inset of Fig.~\ref{Fig_EOS}), which has a nonlinear monotonic decrease
with pressure, the local isothermal compressibility $\kappa_T^0$
demonstrates a significant jump at the pressure $p_c \approx 2000$
Atm. A peculiarity in the vicinity of this value of pressure is also clear observed for
the difference $r_2 - r_1$ presented in Fig.~\ref{Fig_ISOLoc}, although it
is not so obvious in the dependencies $r_1(p)$ and $r_2(p)$. Such result
is a direct evidence that the structural transformations from LDL to
HDL in water at the pressure $p_c \approx 2000$ Atm are related
with the structural rearrangement within the second coordination shell.
Then, the next question arises naturally: Have these structural transformations
an influence on the inherent microscopic dynamics? One of the simplest way to answer on the
question is to consider the vibrational density of states of the system.

\subsection{Vibrational Density of States}

\begin{figure}
\begin{center}
\includegraphics[width=1.1\textwidth]{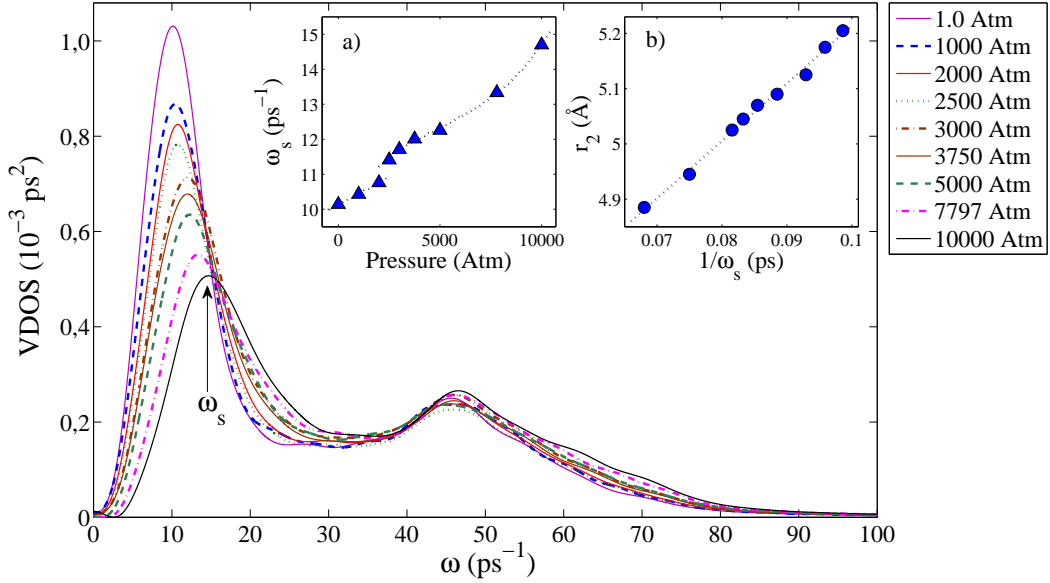}
\end{center}
\caption[kurzform]{\label{Fig_VDOS} (Color online) Main: Vibrational
density of states of liquid water at the temperature $T=290$~K.
Inset: (a) Pressure dependence of the low-frequency peak in VDOS
indicated by arrow on main panel; (b) Radius of the second
coordination shell $r_2$ vs. characteristic time scale associated
with the low-frequency vibrational dynamics, $1/\omega_s$, for
liquid water at the temperature $T=290$~K.}
\end{figure}

The vibrational density of states (VDOS) characterizes the
distribution over frequencies related with the molecular vibrational
degree of freedom~\cite{Kohanoff_1994}:
\begin{equation}
\widetilde{\Phi}(\omega)=\frac{1}{2\pi}\Bigg[
\int_{-\infty}^{\infty} \frac{\big\langle
\vec{\vartheta}(0)\cdot\vec{\vartheta}(t) \big\rangle} {\big\langle
\vec{\vartheta}(0)\cdot\vec{\vartheta}(0) \big\rangle}  e^{i\omega
t} dt \Bigg]^2, \label{Eq_VDOS}
\end{equation}
where $\vec{\vartheta}(t)$ is the velocity of the mass center of a
molecule at the time $t$. Fig. \ref{Fig_VDOS} presents the defined VDOS
of liquid water at the different pressures and the fixed temperature
$T=290$~K. As can be seen, VDOS has a bimodal form for all the cases. It is remarkable that
the high-frequency peak position in VDOS is independent on pressure
whereas the low-frequency peak changes both the intensity and the
position $\omega_s$. As can be see from the pressure dependence of
$\omega_s$ given in inset (a) of Fig.~\ref{Fig_VDOS}, the quantity
$\omega_s$ has a non-monotonic step-like behavior in vicinity of the
characteristic value of pressure $p_c \approx 2000$ Atm. This indicates on the
correlation of the low-frequency molecular vibrational properties
with the structural transformations at the pressure. While
high-frequency mode corresponds to the inherent vibrational
processes of the molecules, the time scale $\tau_s \sim 1/\omega_s$
is related with the vibrational dynamics of the molecular
conglomerates \cite{Heyden_2010,Mokshin_2010}. The immediate
correlation of the low-frequency vibrations with the structural
transformations at the local spatial scales in the range of the
second coordination shell can be directly seen from inset (b) of
Fig.~\ref{Fig_VDOS}.
%%%%%%%%%%%%%%%%%%%%%%%%%%%%%%%%%%%%%%%%%%%%%%%%%%%%%%%%%%%
%% Figure  4
%%%%%%%%%%%%%%%%%%%%%%%%%%%%%%%%%%%%%%%%%%%%%%%%%%%%%%%%%%%
%%%%%%%%%%%%%%%%%%%%%%%%%%%%%%%%%%%%%%%%%%%%%%%%%%%%%%%%%%%%%%%%%
%%%%%%%%%%%%%%%%%%%%%%%%%%%%%%%%%%%%%%%%%%%%%%%%%%%%%%%%%%%%%%%%%
Recently, we have shown \cite{Khusnutdinoff_2011} that the increase of pressure
gives rise the monotonic decrease of the tetrahedral order parameter,
since the spatial structures of the five bonded water molecules changes.
The low-frequency mode can be attributed to the vibrational motions of the separate
molecular tetrahedral complexes.
So, one can concluded that the pressure-induced  changes in the
low-frequency mode is governed by the change of geometry
of the local molecular conglomerates formed by five bonded molecules.

\section{Summary}
Results of this study suggest explicitly the existence of the
structural transformations from low-density state to high density one at a ``stable
liquid region'' of the water phase diagram. Namely, we found for
water system at the temperature $290$~K that the Amoeba model
reveals the transformation in the vicinity of the characteristic value of pressure $p_c \approx
2000$~Atm. It is necessary to note the corresponding structural transformation
from LDL to HDL water is not marked in the
temperature-pressure phase diagrams of water given recently, for
example, in Ref.~\cite{Soper_2008a} (see Fig.~1) and
Ref.~\cite{Stanley_2007} (Fig.~4).

Strikingly, the transformation considered in this work has a lot of similar
features with the \textit{polyamorphic} phase transition from the
low- to high-density amorphous forms of ice, which has been experimentally
observed near 2000 Atm and recognized as a first-order like transition
(see work of O. Mishima et al. \cite{Mishima_2010}).
Furthermore, the transition in amorphous phase has a peculiarity to involve the movement of one
water molecule from the second coordination shell to an interstitial site
in the first coordination shell \cite{Finney_2002a,Bowron_2006}, that is
very similar to LDL/HDL structural transformation considered here.
Nevetheless, in the studied temperature-pressure range of stable
liquid phase, there is no signatures of a phase transition in the thermodynamic
sense between these two forms of water.

The isothermal compressibility $\kappa_T$ of the system decreases
steadily with pressure, while the local extension of the isothermal
compressibility $\kappa_T^0$ as well as other local structural and
vibrational characteristics manifests a step-like behavior at the characteristic
value of pressure $p_c \approx 2000$~Atm. It is found that the LDL/HDL structural
transformation in the vicinity of the pressure $p_c \approx 2000$~Atm has an influence
on the low-frequency vibrational dynamics whereas the high-frequency
molecular dynamics is independent on the pressure. Moreover, there
is a strong correlation between low-frequency vibrational dynamics and the
spatial scale of the second coordination shell.

\section{Acknowledgments}
This work was supported by grant of the Russian Foundation for Basic
Research and the Centre National de la Recherche Scientifique (No. 09-02-91053-CNRS-a).


\begin{thebibliography}{10}

\bibitem{Khusnutdinoff_2011} R.M. Khusnutdinoff, A.V. Mokshin, J. Non-Cryst. Solids \textbf{357}, 1677 (2011).

\bibitem{Soper_2008a} A.K. Soper, Mol. Phys. \textbf{106}, 2053 (2008).

\bibitem{Soper_2000} A.K. Soper and M.A. Ricci, Phys. Rev. Lett. \textbf{84}, 2881 (2000).

\bibitem{Katayama_2010} Y. Katayama, T. Hattori, H. Saitoh et al, Phys. Rev. B \textbf{81}, 014109 (2010).

\bibitem{Davitt_2010} K. Davitt, E. Rolley, F. Caupin, A. Arvengas, and S. Balibar, J. Chem. Phys. \textbf{133}, 174507 (2010).

\bibitem{Krisch_2002} M. Krisch, P. Loubeyre, G. Ruocco et al, Phys. Rev. Lett. \textbf{89} 125502 (2002).

\bibitem{Jansson_2010} H. Jansson, R. Bergman, and J. Swenson, Phys. Rev. Lett. \textbf{104}, 017802 (2010).

\bibitem{Liu_2002} L. Liu, A. Faraone, S.-H. Chen, Phys. Rev. E \textbf{65}, 041506 (2002).

\bibitem{Li_2010} F. Li and J.L. Skinner, J. Chem. Phys. \textbf{133}, 244504 (2010).

\bibitem{Moore_2010} E.B. Moore and V. Molinero, J. Chem. Phys. \textbf{132}, 244504 (2010).

\bibitem{Soper_2008} A.K. Soper and C.J. Benmore, Phys. Rev. Lett. \textbf{101}, 065502 (2008).

\bibitem{Vega_2009} C. Vega, J.L.F. Abascal, M.M. Conde and J.L. Aragones, Faraday Discuss. \textbf{141}, 251 (2009).

\bibitem{Paschek_2005} D. Paschek, Phys. Rev. Lett. \textbf{94}, 217802 (2005).

\bibitem{Salzmann_2009} C.G. Salzmann, P.G. Radaelli, E. Mayer, and J.L. Finney, Phys. Rev. Lett. \textbf{103}, 105701 (2009).

\bibitem{Brovchenko_2008} I. Brovchenko, A. Oleinikova, Chem. Phys. Chem. \textbf{9}, 2660 (2008).

\bibitem{Poole_1992} P.H. Poole, F. Sciortino, U. Essmann and H.E. Stanley, Nature (London) \textbf{360}, 324 (1992).

\bibitem{Mishima_1998} O. Mishima and H.E. Stanley, Nature \textbf{396}, 329 (1998).

\bibitem{Debenedetti_2003} P.G. Debenedetti, J. Phys.: Condens. Matter \textbf{15}, R1669 (2003).

\bibitem{Debenedetti_2003b} P.G. Debenedetti and H.E. Stanley, Phys. Today \textbf{56}, 40 (2003).

\bibitem{Loerting_2006} T. Loerting, C.G. Salzmann, K. Winkel, E. Mayer, Phys. Chem. Chem. Phys. \textbf{8}, 2810 (2006).

\bibitem{Loerting_2001} T. Loerting, C. Salzmann, I. Kohl, E. Mayer, and A. Hallbrucker, Phys.
Chem. Chem. Phys. \textbf{3}, 5355 (2001).

\bibitem{Finney_2002a} J.L. Finney, A. Hallbrucker, I. Kohl, A.K. Soper, D.T. Bowron, Phys. Rev. Lett. \textbf{88}, 225503 (2002).

\bibitem{Finney_2002b} J.L. Finney, D.T. Bowron, A.K. Soper, T. Loerting, E. Mayer, and A. Hallbrucker,
Phys. Rev. Lett. \textbf{89}, 205503 (2002).

\bibitem{Loerting_PRL2006} T. Loerting, W. Schustereder, K. Winkel, C. Salzmann, I. Kohl, E. Mayer,
Phys. Rev. Lett. \textbf{96}, 025701 (2006).

\bibitem{Winkel_2008} K. Winkel, M.S. Elsaesser, E. Mayer, T. Loerting, J. Chem. Phys. \textbf{128}, 044510 (2008).

\bibitem{Stanley_2007} H.E. Stanley, P. Kumar, L. Xu et al, Physica A \textbf{386}, 729 (2007).

\bibitem{Mishima_2010} O. Mishima, J. Chem. Phys. \textbf{133}, 144503 (2010).

\bibitem{Saitta_2003} A.M. Saitta and F. Datchi, Phys. Rev. E \textbf{67}, 020201(R) (2003).

\bibitem{Li_2005} F.Li, Q. Cui, Z. He, T. Cui et al, J. Chem. Phys. \textbf{123}, 174511 (2005).

\bibitem{Mokshin_2005} A.V. Mokshin, R.M. Yulmetyev, and P. H\"anggi, Phys. Rev. Lett. \textbf{95}, 200601 (2005).

\bibitem{Soper_2000b} A.K. Soper, Chem. Phys. \textbf{258}, 121 (2000).

\bibitem{Ren_2003} P. Ren and J.W. Ponder, J. Phys. Chem. B \textbf{107}, 5933 (2003).

\bibitem{Ren_2004} P. Ren and J.W. Ponder, J. Phys. Chem. B \textbf{108}, 13427 (2004).

\bibitem{Thole_1981} B.T. Thole, Chem. Phys. \textbf{59}, 341 (1981).

\bibitem{Ponder_2003} J.W. Ponder, TINKER: \textit{Software Tools for Molecular Design}; 4.1
ed.: Washington University School of Medicine: Saint Louis, (2003).

\bibitem{Allen_Tildesley} M.P. Allen and D.J. Tildesley, \textit{Computer Simulation of Liquids}, Clarendon Press, Oxford (1987).

\bibitem{Berendsen_1984} H.J.C. Berendsen, J.P.M. Postma, W.F. van Gunsteren, A. DiNola, J.R. Haak, J. Chem. Phys. \textbf{81}, 3684 (1984).

\bibitem{Yan_2007} Zh. Yan, S.V. Buldyrev, P. Kumar, N. Giovambattista, P.G. Debenedetti, H.E. Stanley,
Phys. Rev. E \textbf{76}, 051201 (2007).

\bibitem{Kohanoff_1994} J. Kohanoff, Comp. Mat. Sci. \textbf{2}, 221 (1994).

\bibitem{Heyden_2010} M. Heyden, J. Sun, S. Funkner, G. Mathias, H. Forbert, M. Havenith, D. Marx, Proc. Natl. Acad. Sci. \textbf{107}, 12068 (2010).

\bibitem{Mokshin_2010} A.V. Mokshin and J.-L. Barrat, Phys. Rev. E \textbf{82}, 021505 (2010).

\bibitem{Bowron_2006} D.T. Bowron, J.L. Finney, A. Hallbrucker, I. Kohl, T. Loerting, E. Mayer, A.K. Soper, J. Chem. Phys. \textbf{125}, 194502 (2006).

\end{thebibliography}
\end{document}